\documentstyle[twoside,epsf]{article}

\catcode`\@=11
\long\def\@makefntext#1{
\protect\noindent \hbox to 3.2pt {\hskip-.9pt  
$^{{\eightrm\@thefnmark}}$\hfil}#1\hfill}               %CAN BE USED 

\def\@makefnmark{\hbox to 0pt{$^{\@thefnmark}$\hss}}    %ORIGINAL 
        
\def\ps@myheadings{\let\@mkboth\@gobbletwo
\def\@oddhead{\hbox{}
\rightmark\hfil\eightrm\thepage}   
\def\@oddfoot{}\def\@evenhead{\eightrm\thepage\hfil
\leftmark\hbox{}}\def\@evenfoot{}
\def\sectionmark##1{}\def\subsectionmark##1{}}

\oddsidemargin=\evensidemargin
\newcounter{sectionc}\newcounter{subsectionc}\newcounter{subsubsectionc}
\renewcommand{\section}[1] {\vspace{12pt}\addtocounter{sectionc}{1} 
\setcounter{subsectionc}{0}\setcounter{subsubsectionc}{0}\noindent 
        {\tenbf\thesectionc. #1}\par\vspace{5pt}}
\renewcommand{\subsection}[1] {\vspace{12pt}\addtocounter{subsectionc}{1} 
        \setcounter{subsubsectionc}{0}\noindent 
        {\bf\thesectionc.\thesubsectionc. {\kern1pt \bfit #1}}\par\vspace{5pt}}
\renewcommand{\subsubsection}[1] {\vspace{12pt}\addtocounter{subsubsectionc}{1}
        \noindent{\tenrm\thesectionc.\thesubsectionc.\thesubsubsectionc.
        {\kern1pt \tenit #1}}\par\vspace{5pt}}
\newcommand{\nonumsection}[1] {\vspace{12pt}\noindent{\tenbf #1}
        \par\vspace{5pt}}

\newcounter{appendixc}
\newcounter{subappendixc}[appendixc]
\newcounter{subsubappendixc}[subappendixc]
\renewcommand{\thesubappendixc}{\Alph{appendixc}.\arabic{subappendixc}}
\renewcommand{\thesubsubappendixc}
        {\Alph{appendixc}.\arabic{subappendixc}.\arabic{subsubappendixc}}

\renewcommand{\appendix}[1] {\vspace{12pt}
        \refstepcounter{appendixc}
        \setcounter{figure}{0}
        \setcounter{table}{0}
        \setcounter{lemma}{0}
        \setcounter{theorem}{0}
        \setcounter{corollary}{0}
        \setcounter{definition}{0}
        \setcounter{equation}{0}
        \renewcommand{\thefigure}{\Alph{appendixc}.\arabic{figure}}
        \renewcommand{\thetable}{\Alph{appendixc}.\arabic{table}}
        \renewcommand{\theappendixc}{\Alph{appendixc}}
        \renewcommand{\thelemma}{\Alph{appendixc}.\arabic{lemma}}
        \renewcommand{\thetheorem}{\Alph{appendixc}.\arabic{theorem}}
        \renewcommand{\thedefinition}{\Alph{appendixc}.\arabic{definition}}
        \renewcommand{\thecorollary}{\Alph{appendixc}.\arabic{corollary}}
        \renewcommand{\theequation}{\Alph{appendixc}.\arabic{equation}}
        \noindent{\tenbf Appendix \theappendixc #1}\par\vspace{5pt}}
\newcommand{\subappendix}[1] {\vspace{12pt}
        \refstepcounter{subappendixc}
        \noindent{\bf Appendix \thesubappendixc. {\kern1pt \bfit #1}}
        \par\vspace{5pt}}
\newcommand{\subsubappendix}[1] {\vspace{12pt}
        \refstepcounter{subsubappendixc}
        \noindent{\rm Appendix \thesubsubappendixc. {\kern1pt \tenit #1}}
        \par\vspace{5pt}}

\newcommand{\textlineskip}{\baselineskip=13pt}
\newcommand{\smalllineskip}{\baselineskip=10pt}

\def\eightcirc{
\begin{picture}(0,0)
\put(4.4,1.8){\circle{6.5}}
\end{picture}}
\def\eightcopyright{\eightcirc\kern2.7pt\hbox{\eightrm c}} 

\newcommand{\copyrightheading}[1]
        {\vspace*{-2.5cm}\smalllineskip{\flushleft
        {\footnotesize International Journal of Modern Physics A, #1}\\
        {\footnotesize $\eightcopyright$\, World Scientific Publishing
         Company}\\
         }}

\def\abstracts#1#2#3{{
        \centering{\begin{minipage}{4.5in}\baselineskip=10pt\footnotesize
        \parindent=0pt #1\par 
        \parindent=15pt #2\par
        \parindent=15pt #3
        \end{minipage}}\par}} 

\renewenvironment{thebibliography}[1]
        {\frenchspacing
         \ninerm\baselineskip=11pt
         \begin{list}{\arabic{enumi}.}
        {\usecounter{enumi}\setlength{\parsep}{0pt}
         \setlength{\leftmargin 12.7pt}{\rightmargin 0pt} %FOR 1--9 ITEMS
         \setlength{\itemsep}{0pt} \settowidth
        {\labelwidth}{#1.}\sloppy}}{\end{list}}

\newcounter{itemlistc}
\newcounter{romanlistc}
\newcounter{alphlistc}
\newcounter{arabiclistc}

\newcommand{\fcaption}[1]{
        \refstepcounter{figure}
        \setbox\@tempboxa = \hbox{\footnotesize Fig.~\thefigure. #1}
        \ifdim \wd\@tempboxa > 5in
           {\begin{center}
        \parbox{5in}{\footnotesize\smalllineskip Fig.~\thefigure. #1}
            \end{center}}
        \else
             {\begin{center}
             {\footnotesize Fig.~\thefigure. #1}
              \end{center}}
        \fi}

\newcommand{\tcaption}[1]{
        \refstepcounter{table}
        \setbox\@tempboxa = \hbox{\footnotesize Table~\thetable. #1}
        \ifdim \wd\@tempboxa > 5in
           {\begin{center}
        \parbox{5in}{\footnotesize\smalllineskip Table~\thetable. #1}
            \end{center}}
        \else
             {\begin{center}
             {\footnotesize Table~\thetable. #1}
              \end{center}}
        \fi}

\def\@citex[#1]#2{\if@filesw\immediate\write\@auxout
        {\string\citation{#2}}\fi
\def\@citea{}\@cite{\@for\@citeb:=#2\do
        {\@citea\def\@citea{,}\@ifundefined
        {b@\@citeb}{{\bf ?}\@warning
        {Citation `\@citeb' on page \thepage \space undefined}}
        {\csname b@\@citeb\endcsname}}}{#1}}

\newif\if@cghi
\def\cite{\@cghitrue\@ifnextchar [{\@tempswatrue
        \@citex}{\@tempswafalse\@citex[]}}
\def\citelow{\@cghifalse\@ifnextchar [{\@tempswatrue
        \@citex}{\@tempswafalse\@citex[]}}
\def\@cite#1#2{{$\null^{#1}$\if@tempswa\typeout
        {IJCGA warning: optional citation argument 
        ignored: `#2'} \fi}}

\def\pmb#1{\setbox0=\hbox{#1}
        \kern-.025em\copy0\kern-\wd0
        \kern.05em\copy0\kern-\wd0
        \kern-.025em\raise.0433em\box0}

\def\fnt#1#2{\footnotetext{\kern-.3em
        {$^{\mbox{\scriptsize #1}}$}{#2}}}

\def\fpage#1{\begingroup
\voffset=.3in
\thispagestyle{empty}\begin{table}[b]\centerline{\footnotesize #1}
        \end{table}\endgroup}

\def\runninghead#1#2{\pagestyle{myheadings}
\markboth{{\protect\footnotesize\it{\quad #1}}\hfill}
{\hfill{\protect\footnotesize\it{#2\quad}}}}
\font\tenrm=cmr10
\font\tenit=cmti10 
\font\tenbf=cmbx10
\font\bfit=cmbxti10 at 10pt
\font\ninerm=cmr9

\font\eightrm=cmr8

\def\qed{\hbox{${\vcenter{\vbox{                        %HOLLOW SQUARE
   \hrule height 0.4pt\hbox{\vrule width 0.4pt height 6pt
   \kern5pt\vrule width 0.4pt}\hrule height 0.4pt}}}$}}

\begin{document}

\runninghead{Semi-Contained Neutrino Events in MACRO}{Semi-Contained Neutrino Events in MACRO}

\normalsize\textlineskip
\thispagestyle{empty}
\setcounter{page}{1}

\copyrightheading{}                     %{Vol. 0, No. 0 (1993) 000--000}

\vspace*{0.88truein}

\fpage{1}
\centerline{\bf SEMI-CONTAINED NEUTRINO EVENTS IN MACRO}
\vspace*{0.37truein}
\centerline{\footnotesize R. NOLTY}
\vspace*{0.015truein}
\centerline{\footnotesize\it California Institute of Technology}
\baselineskip=10pt
\centerline{\footnotesize\it Pasadena, California 91125}
%\vspace*{0.225truein}
%\publisher{(received date)}{(revised date)}

\vspace*{0.21truein} \abstracts{Updated results are presented of
  low-energy ($\overline{E_\nu} \sim 5\ GeV$) neutrino interactions
  observed by the MACRO detector.  Two analyses (of different topologies) are
  presented; individually, and especially in their ratio, they are
  inconsistent with no oscillations and consistent with maximal mixing
  at  $\Delta m^2$ of a few times $10^{-3}$.}{}{}

%\textlineskip                  %) USE THIS MEASUREMENT WHEN THERE IS
%\vspace*{12pt}                 %) NO SECTION HEADING

\vspace*{1pt}\textlineskip      %) USE THIS MEASUREMENT WHEN THERE IS
\section{Introduction}          %) A SECTION HEADING
\vspace*{-0.5pt}

\noindent
Recent measurements of atmospheric neutrino flux by the
Super-Kamiokande\cite{SuperK}, Soudan\cite{Soudan} and
MACRO\cite{doug} experiments all suggest oscillations with $\Delta
m^2$ a few times $10^{-3}$ and $sin^2 2\theta \sim 1$.  The MACRO
analysis has recently been extended\cite{lownu} to event topologies
that probe lower neutrino energies, and the results are updated here.

\begin{figure}[htbp]
\vspace*{13pt}
\centerline{\vbox{\hrule width 5cm height0.001pt}}
\centerline{\epsfxsize 2 truein \epsfbox{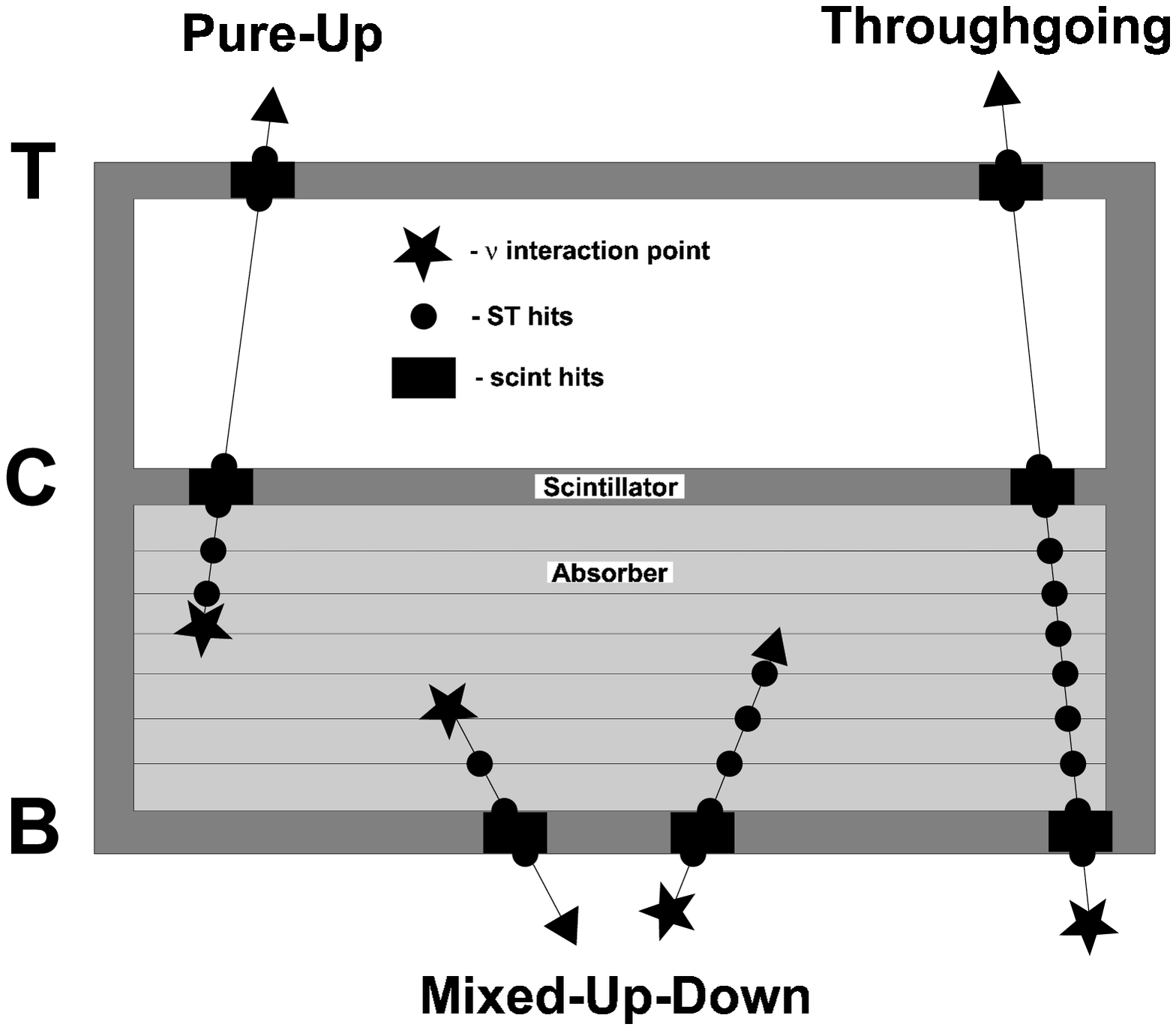}
  \epsfxsize 2 truein \epsfbox{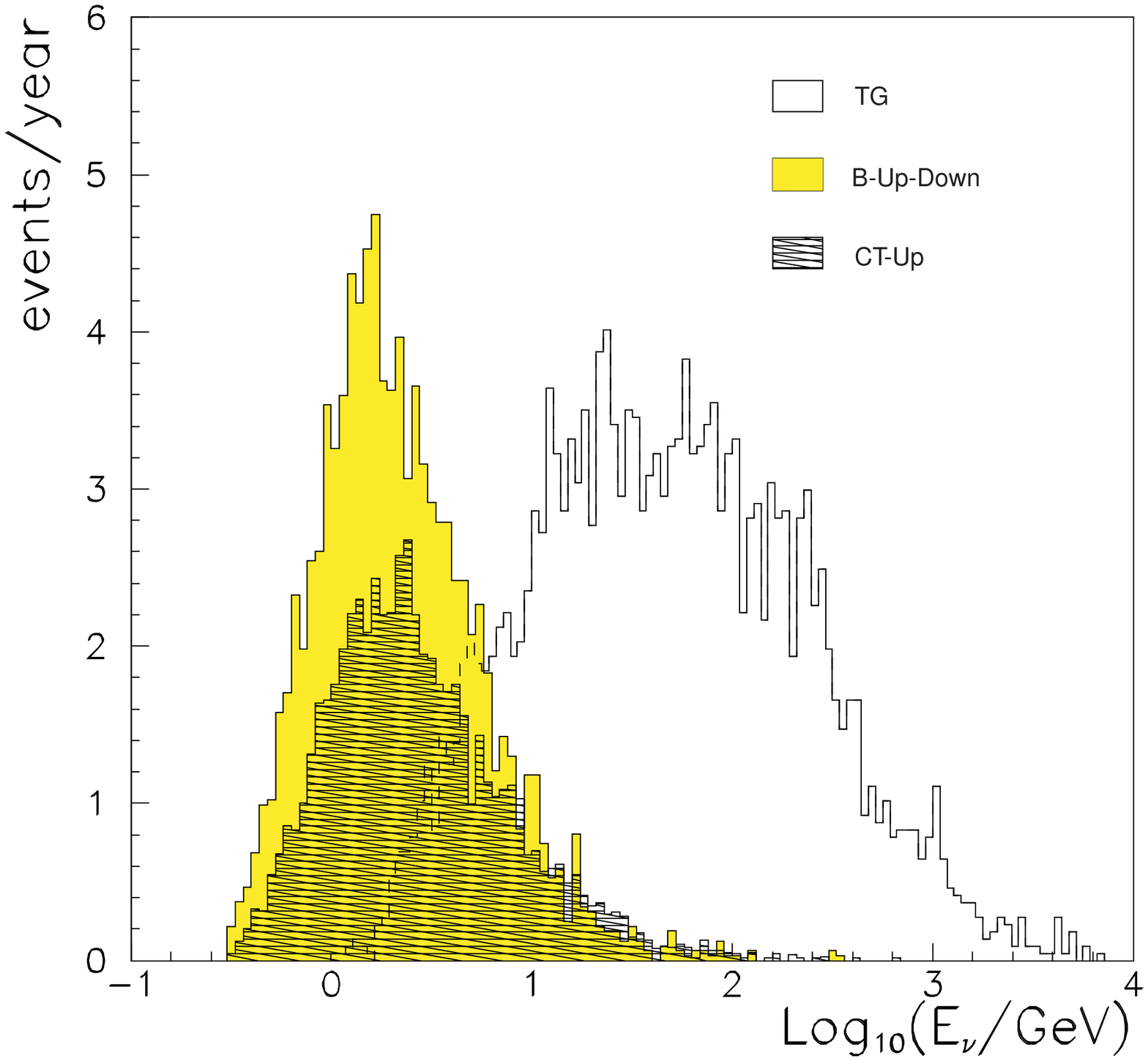}}
\centerline{\vbox{\hrule width 5cm height0.001pt}}
\vspace*{13pt}
\fcaption{\protect\label{fig:sctop}Event topologies of
  neutrino-induced events in MACRO.}
\end{figure}

MACRO\cite{MACROTech} is a large detector located deep underground at
the Gran Sasso laboratory in Italy.  The active detector elements are
layers of liquid scintillator and layers of streamer tubes (with wire
and strip views) with a pitch of 3 cm.  Most neutrino interactions
take place in the massive bottom half of the detector which is filled
with crushed rock absorber.  The interior of the upper portion of the
detector is hollow.  (See Figure~\ref{fig:sctop}.)

\pagebreak
\section{The Analyses}

\noindent
Because of its large granularity, MACRO is sensitive only to charged
current $\nu_\mu$ interactions producing a muon that travels at least
tens of cm.  This talk examines two topologies of lower-energy
neutrino-induced muons.  The two analyses have very similar parent
neutrino energy distributions, with $\overline{E_\nu} \sim 5\ GeV$.
Upward contained-vertex events, in which the muon strikes two layers
of scintillator and exits the detector, are labeled Pure-Up in
Figure~\ref{fig:sctop}.  Downward contained-vertex events and upward
stopping events from below both hit only the bottom scintillator layer
and have a few associated colinear streamer tube hits.  In MACRO the
direction (up or down) of these events cannot be determined and they
are merged into an analysis labeled Mixed-Up-Down.

%%% bob %%% \pagebreak

%%% bob %%% \textheight=7.8truein
%%% bob %%% \setcounter{footnote}{0}
%%% bob %%% \renewcommand{\thefootnote}{\alph{footnote}}

Predicted event rates are made using a Monte Carlo calculation
combining a neutrino flux model, a neutrino cross section model, and
detailed simulation of the detector geometry and response.  We chose
the Bartol flux calculation\cite{Bartol} including geomagnetic
effects, and the Lipari cross section model\cite{Lipari} which
includes quasi-elastic and resonant scattering in addition to deep
inelastic scattering.  The deep inelastic portion was calculated using
the GRV-LO-94 parton distribution functions\cite{GRV}.

\section{Results}
\begin{figure}[htbp]
\vspace*{13pt}
\centerline{\vbox{\hrule width 5cm height0.001pt}}
\centerline{\epsfxsize 3 truein \epsfbox{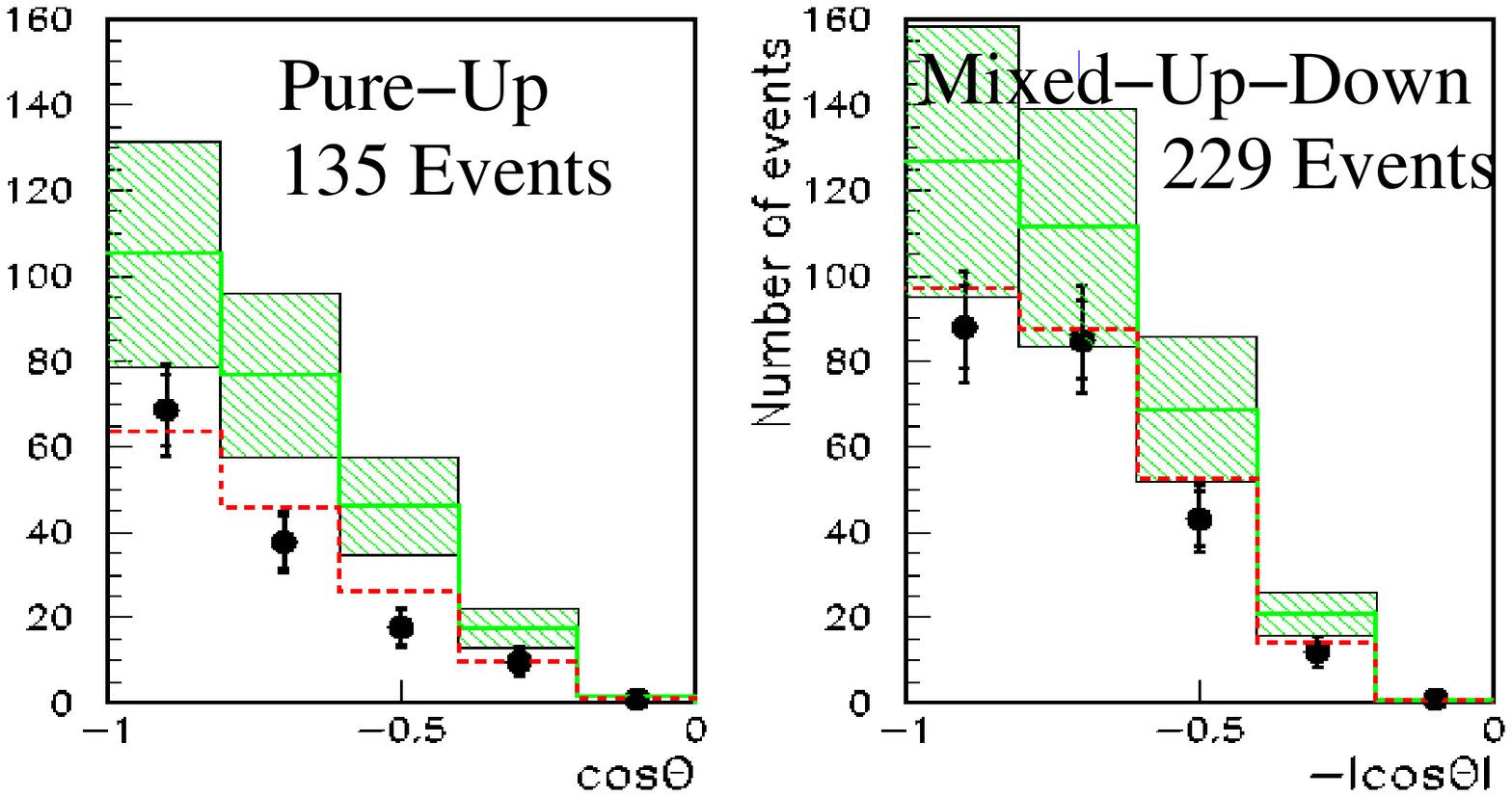}
  \epsfxsize 1.6 truein \epsfbox{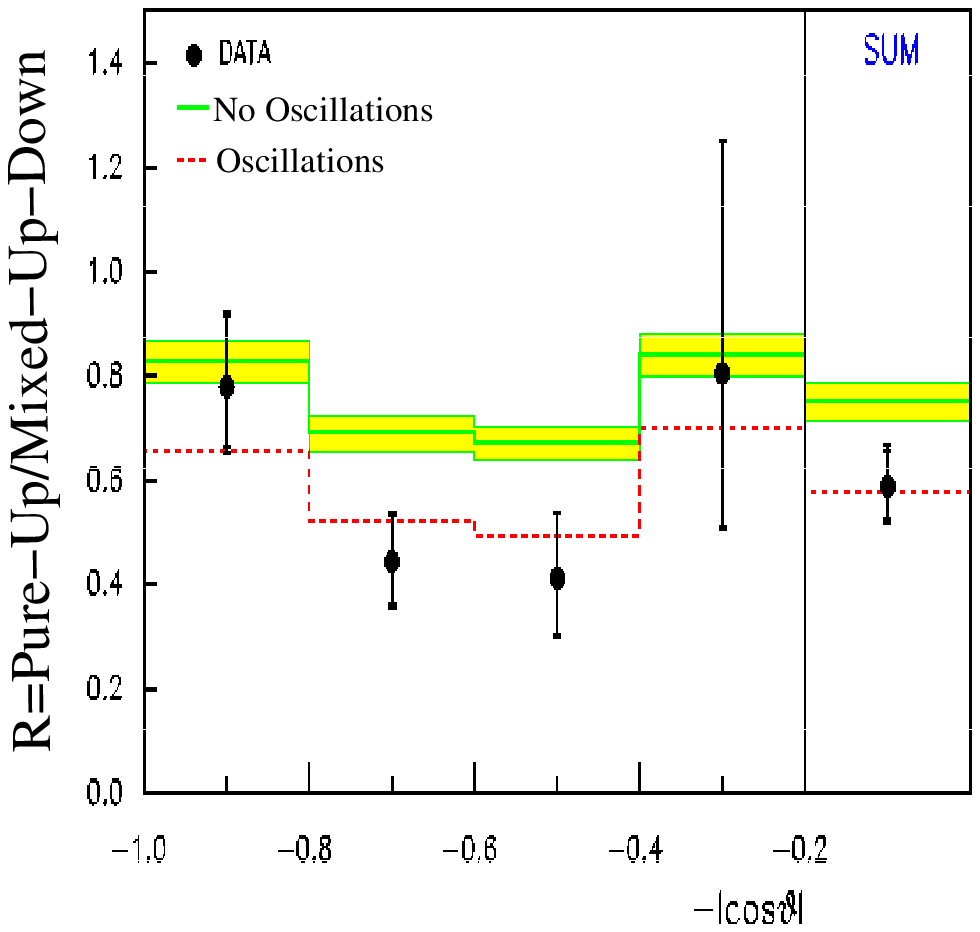}}
\centerline{\vbox{\hrule width 5cm height0.001pt}}
\vspace*{13pt}
\fcaption{\protect\label{fig:results}Zenith distributions of events
  selected in the two low-energy analyses.  The shaded region gives the
  prediction (with uncertainties) of a no-oscillations Monte Carlo.
  The dashed line gives  the prediction for $\Delta m^2 = 2.5 \times
  10^{-3}$ and $sin^2  2\theta = 1$.  The third figure gives the ratio
  of the two analyses.}
\end{figure}

\noindent
Here results are updated through March, 2000 for a total of 5.1 live
years, an increment of 25\% over our last published
result\cite{lownu}.  Angular distributions, compared to
no-oscillations and oscillated predictions, are given in
Figure~\ref{fig:results}.  Uncertainties on the neutrino flux (20\%)
and cross section (15\%) lead to a large theoretical uncertainty in
the predicted rates.

Integrating over all zenith angle bins and forming the ratio of
observed to expected events, we find $
R_{Pure-Up} = 0.55 \pm 0.04_{stat} \pm 0.06_{sys} \pm 0.14_{theor}
$ and $
R_{Mixed-Up-Down} = 0.70 \pm 0.04_{stat} \pm 0.07_{sys} \pm 0.18_{theor}
$.
$R_{Pure-Up}$ could be a statistical fluctuation from the
no-oscillations model with probability 4.3\%.  For $R_{Mixed-Up-Down}$
the probability is 12\%.  It is the theoretical uncertainty on the
flux and cross section normalizations that makes these numbers so
large.

\begin{figure}[htbp]
\vspace*{13pt}
\centerline{\vbox{\hrule width 5cm height0.001pt}}
\centerline{\epsfxsize 1.7 truein \epsfbox{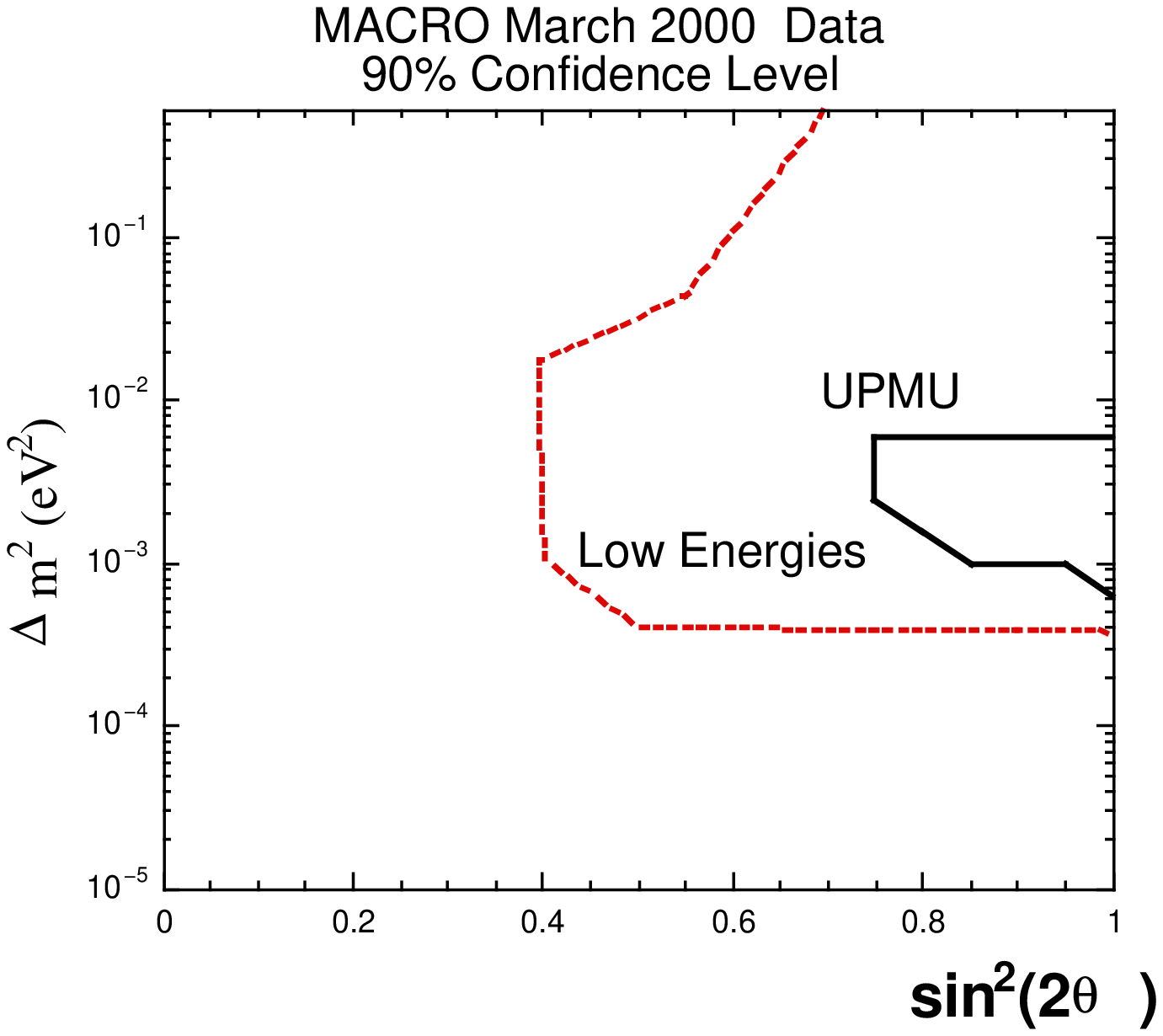}}
\centerline{\vbox{\hrule width 5cm height0.001pt}}
\vspace*{13pt}
\fcaption{\protect\label{fig:exclusion}Region of parameter space
  excluded by this analysis at 90\% confidence level, which is
  compatible with the more precise result from MACRO's high energy
  analysis\cite{doug} (labeled UPMU).}
\end{figure}

We can reduce the uncertainties by considering the two analyses
simultaneously rather than independently.  For example, if we consider
one of the measurements to fix the normalization at a level far below
the calculated normalization, we find that the other measurement is
incompatible with that normalization.  To put it in different
language, when we form the ratio of the two analyses, most theoretical
error and some systematic error cancels.  This comes at the expense of
a greater statistical uncertainty, because we are dividing two
uncertain numbers by each other.  The results (also shown in
Figure~\ref{fig:results}) are $ R_{Data} = 0.59 \pm 0.07_{stat} $;$
R_{Expected} = 0.75 \pm 0.04_{sys} \pm 0.04_{theor} $.  The
probability of attaining this result due to statistical fluctuations
is only 2.7\%.  Combining the ratio and the individual measurements we
may deduce the exclusion region in oscillations parameter space shown
in Figure~\ref{fig:exclusion}.

\nonumsection{References}

\end{document}